\documentclass{article}
\newfont{\feff}{cmti10}
\usepackage{amsbsy}
\usepackage[intlimits]{amsmath}
\usepackage{amsfonts,amssymb}
\DeclareSymbolFontAlphabet{\mathbb}{AMSb}
\usepackage{float,subfigure}
\usepackage[bf]{caption2}
\setcaptionmargin{0.5in}
\usepackage{fancyheadings,fancybox}
\usepackage{url}
\usepackage{lscape,afterpage}
\usepackage{xspace}
\usepackage{graphicx}
\usepackage{subfigure}
\usepackage{natbib}

\providecommand\bnabla{\boldsymbol{\nabla}}
\providecommand\bcdot{\boldsymbol{\cdot}}
\newcommand\Real{\mbox{Re}} 
 
\newcommand\Rey{\mbox{\textit{Re}}}

\title{Lattice Boltzmann Simulation of Electromechanical Resonators in Gaseous Media}

\title{Lattice Boltzmann Simulation of High-Frequency Flows: Electromechanical Resonators in Gaseous Media}
\author{Carlos Colosqui$^1$ \thanks{Present address: Department of Chemical Engineering,
Princeton University, Princeton, NJ 08544, USA} , Devrez M. Karabacak$^2$, Kamil L. Ekinci$^1$ {\small AND} Victor Yakhot$^1$\\ 
$^1$Department of Mechanical Engineering, Boston University, Boston, MA 02215, USA\\
$^2$IMEC Holst Centre, Eindhoven, 5605 KN, The Netherlands\\
}

\date{April 2009}

\begin{document}

\maketitle

\begin{abstract}
In this work, we employ a kinetic theory based approach to predict the hydrodynamic forces on electromechanical resonators operating in gaseous media.
 Using the Boltzmann-BGK equation, we investigate the influence of the resonator geometry on the fluid resistance in the entire range of nondimensional frequency variation $0\le\tau\omega\le\infty$; here the fluid relaxation time $\tau=\mu/p$ is determined by the gas viscosity $\mu$ and pressure $p$ at thermodynamic equilibrium, and $\omega$ is the (angular) oscillation frequency. Our results support the experimentally observed transition from viscous to viscoelastic flow in simple gases at $\tau\omega\approx1$. They are also in remarkable agreement with the measured geometric effects in resonators in a broad linear dimension, frequency, and pressure range.
\end{abstract}

\section{Introduction}

Electromechanical resonators with linear dimensions in the nanometer
to micrometer scales are being developed for technological
applications and fundamental research. One of the most important
attributes of these nano/microelectromechanical systems (N/MEMS)
resonators is that they have very small intrinsic dissipation of
energy, quantified by their high quality factors $Q_o={\cal
O}(10^2-10^4)$. N/MEMS resonators are thus ultrasensitive to
external perturbations enabling important technologies such as
atomic force microscopy (AFM) \cite[]{Binnig} and
bio-chemical sensing \cite[]{Ekinci2004}.

Some of the most promising applications of N/MEMS, however, require
their immersion in fluid media (e.g., air mixtures or biological
fluids), where fluid-device interaction can significantly degrade
the overall sensitivity \cite[]{Bhiladvala,Sader,Paul}. Numerous
efforts are currently underway to overcome this difficulty and
develop future N/MEMS for promising nanotechnological and biomedical
applications. Unquestionably, the flows generated by N/MEMS demand a
novel understanding of fluid mechanics at increasingly smaller time
and length scales. Conversely, experimental characterization and
numerical analysis of fluid-immersed N/MEMS resonators provide an
invaluable opportunity to advance knowledge in new areas of fluid
mechanics such as high-frequency nanofluidics. Recent work on
high-frequency oscillating flows
\cite[]{Karabacak2007,Yakhot,Colosqui2009} reports a viscoelastic
transition in simple gases at sufficiently large values of the
nondimensional frequency $\tau\omega$. Here, $\tau=\mu/p$ is the
relaxation time in terms of the pressure $p$ and viscosity $\mu$ of
the gas at equilibrium; $\omega$ is the oscillation frequency. Such
remarkable phenomenon is beyond the reach of classical (Newtonian)
fluid mechanics, which is only valid for $\tau\omega\ll 1$. The
viscoelastic transition is accompanied by a substantial attenuation
of the energy dissipated by the fluid and a subsequent improvement
in the performance of the fluid-immersed device.

In the present study, we solve the Boltzmann-BGK equation of kinetic
theory via appropriate numerical procedures
\cite[]{Colosqui2009,Shan2006} in order to predict fluidic effects
(e.g., damping force and energy dissipation) for specific
geometries. The validity of the kinetic methods is not constrained
to Newtonian flow; thus, our models yield excellent agreement with
experimental measurements on different resonators over a wide range
of frequency and pressure variation.


\section{Electromechanical resonators}
\label{resonators}

\begin{figure}
\centerline{\includegraphics[angle=0,scale=0.3]{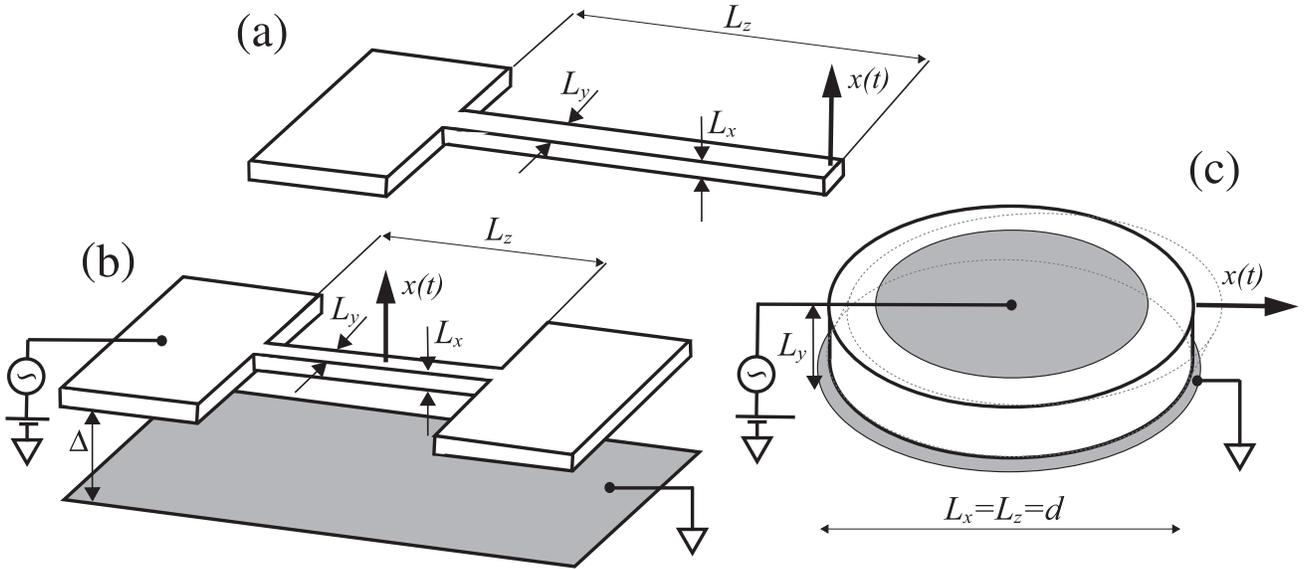}}
\caption{(a) Cantilever Beam (b) Doubly-Clamped Beam (c) Quartz Crystal}
\label{fig:NEMS}
\end{figure}

Illustrated in figures~\ref{fig:NEMS}(a)-\ref{fig:NEMS}(b) are the
first class of studied resonators in the form of cantilever and
doubly-clamped beams ($L_z \gg L_x \sim L_y$). Harmonic motion in
the beams can be induced through the application of periodic
electrostatic, photothermal, or inertial forces. The beams are
driven around their fundamental and first harmonic out-of-plane
flexural resonances while optical techniques are used to determine
the resonant response \cite[]{Kouh}. The doubly-clamped beams are
suspended above a stationary substrate at a mean height $\Delta
\simeq 400$ nm; thus, the presence of the substrate has no
significant effect on the fluidic damping (e.g. via squeeze-film
damping) \cite[]{Karabacak2007}. We also study a macroscopic quartz
crystal resonator \cite[]{Ekinci2008}. The studied resonator [see figure~\ref{fig:NEMS}(c)] is in the form of a thin crystal disk ($L_x=L_z \gg L_y$) connected to electrodes so that its resonances in
thickness-shear modes can be electrically excited and detected by
piezoelectric effects. In all the measurements, the resonance
amplitudes of the beams and crystal resonators are kept extremely
small.

Specific dimensions and (vacuum) characteristics, such as resonance frequency $\omega_o$, quality factor $Q_o$, and surface to modal mass ratio $S/m_o$, of four studied resonators are listed in table~\ref{tab:nems}. The size of the devices vary from sub-micron to millimeters while their resonance frequencies are in the range of kilohertz to megahertz. Experimental analysis of the four resonators in table~\ref{tab:nems} is performed with the devices operating in purified nitrogen at room temperature $T\simeq 300$ K; the pressure is gradually varied from low-vacuum to atmospheric pressure, $0.1\le p \le 1000$ Torr. As the pressure is varied and given that $\tau=\mu/p$, the resulting flows cover a wide range of dimensionless frequency variation $0.001\le\tau\omega\le 10$ \cite[]{Karabacak2007}.

\begin{table}
\begin{center}
\begin{tabular}{@{}lccccc@{}}
Class  & Dimensions & ${\omega_o}/{2\pi}$ & $Q_o$ & $S/m_o$ & $AR$ \\
{\small($\star$:~first harmonic)} &  ($\mu$m) & (MHz) & ~ &  (m$^{2}$/kg) & ~ \\
\hline

Cantilever Beam & $L_x=2.0$~ $L_y=53$~~ $L_z=460$
& 0.078 & 8320 & 688 &  26 \\[0.5ex]

Cantilever Beam$^{\star}$ & $L_x=3.6$~ $L_y=36$~~ $L_z=125$ &
1.97  & 3520 & 405 & 10 \\[0.5ex]

Doubly-Clamped Beam & $L_x=0.2$~ $L_y=0.23$ $L_z=9.6$ &
24.2 & 415  & 8380 & 1  \\[0.5ex]

Quartz Crystal Disk & $L_x\!=\!L_z\!=\!d \simeq 10000$ $L_y\simeq 160$  &
32.7 & 40755  & 6.29 & 0  \\
\end{tabular}
\label{tab:nems}
\caption{Electromechanical Resonators}
\end{center}
\end{table}

{\it Resonators immersed in a fluid}. The dynamics of an
electromechanical resonator immersed in a fluid can be studied by
means of a one-dimensional harmonic oscillator approximation
\cite[]{Cleland}
\begin{equation}
 m_o (x_{tt}+ \gamma_o~x_{t}+ \omega^2_o x)= F(t)+F_f(t)
\label{eq:nems_x}
\end{equation}
where $m_o$ is the effective mass corresponding to the vibrational mode
\cite[]{Karabacak2007,Cleland}, $\gamma_o$ is the (structural)
damping coefficient and $\omega_o$ is the resonance frequency of the
device in vacuum. The load on the oscillator is produced by the
driving force $F(t)=\Real\{F(\omega) e^{-i\omega t}\}$
along with a fluid resistance $F_f(t)=-m_o
(\gamma_f~{x_t}+\beta_f~x_{tt})$ which has both dissipative and
inertial components \cite[]{Landau}. The oscillation amplitude has the
general form $x(t)=\Real\{x(\omega)e^{-i(\omega t+\phi)}\}$ and from equation (\ref{eq:nems_x}) the frequency response is 
\begin{equation}
x(\omega)= \frac{F_0}{m_o}~\frac{1}
{\left[\omega_o^2-\omega^2(1+\beta_f)-i\omega (\gamma_o+\gamma_f)\right]},
\label{eq:nems_uw}
\end{equation}
with $F_0\equiv F(\omega) e^{i \phi}$ being the effective
force amplitude. Equation (\ref{eq:nems_uw}) includes fluidic
effects through the fluidic inertia, or fluid-added mass, $\beta_f$
and fluidic damping $\gamma_f$. Experimental values of the added
mass and fluidic dissipation are respectively obtained from the
frequency shift and the broadening of the Lorentzian frequency
response (\ref{eq:nems_uw}) \cite[]{Karabacak2007}. In this study, the
fluid-added mass is very small $\beta_f \ll 1$ and so is the
measured shift in the resonance frequency: $\Delta\omega_o/\omega_o
\approx \beta_f/2$. The device quality factor in the fluid thus
becomes $Q \approx \omega_o /(\gamma_o+\gamma_f$).

\section{Hydrodynamics of high-frequency flows}
\label{background} Similar  to previous work \cite[]{Bhiladvala,Sader,Paul}, the present analysis is valid when gradients along the spanwise direction of the oscillating body are negligible so that the flow is considered two-dimensional.
This assumption holds for a slender beam ($L_z \gg L_x \sim L_y$)
with uniform rectangular cross-section, or a thin disk ($L_x=L_z=d
\gg  L_y$), where the aspect ratio $AR\!=\!L_y/L_x$ (see table~\ref{tab:nems}) becomes the geometric parameter characterizing the dominant chordwise flow. Other dynamically relevant flow parameters are the Mach number $M=U_o/c_s$ and the Reynolds number $\Rey=\rho U_o L_y/\mu$ determined by the fluid velocity amplitude $U_o=|\omega x(\omega)|$, the speed of sound $c_s\simeq 298$ m s$^{-1}$, and molecular viscosity $\mu\simeq1.78\times 10^{-5}$ kg m$^{-1}$ s$^{-1}$ of nitrogen at room temperature. Given the tiny oscillation amplitude of the resonators, the resulting Mach and Reynolds number are extremely small, $M\sim \Rey <0.001$ in all operation conditions. Consequently, the generated flow can be
assumed laminar, nearly incompressible, and isothermal.

The common approach hitherto encountered in the literature \cite[see][]{Bhiladvala,Sader,Paul} for the theoretical and/or numerical determination of fluidic effects on resonators is based on the Navier-Stokes equations for Newtonian fluid flow. However, a fundamental assumption for the applicability of classical
Navier-Stokes equations is that hydrodynamic scales are much larger
than their kinetic counterparts, i.e. $\tau\omega\ll1$. Previous
work \cite[]{Karabacak2007,Yakhot,Colosqui2009,Ekinci2008}
on high-frequency oscillating flows has established that the
non-dimensional frequency $\tau\omega$ determines qualitatively
different behavior: purely viscous (Newtonian) for $\tau\omega=0$,
viscoelastic (transitional) for $0<\tau\omega<\infty$, and purely
elastic (free-molecular) for $\tau\omega\to \infty$. While viscous
flows are accurately described by Newtonian hydrodynamic equations,
an {\it extended} hydrodynamic description accounting for kinetic
(non-equilibrium) phenomena is required at sufficiently large
$\tau\omega$ where non-Newtonian behavior is observed
\cite[see][]{Yakhot,Colosqui2009}.

\subsection{Newtonian hydrodynamics}
\label{sec:high-newtonian} 
In the Newtonian regime $\tau\omega\ll
1$, the flow around a body oscillating with very small amplitude
$|x(\omega)|\ll L_x$ is governed by the linearized Navier-Stokes
(NS) equations for incompressible flow \cite[]{Landau}:
\begin{equation}
\bnabla \bcdot {\bf u} = 0, ~~~~~ \frac{\partial {\bf u}}{\partial t}=
\nu \nabla^2 {\bf u} - \frac{1}{\rho}\bnabla p. \label{eq:nems_N-S}
\end{equation}
After adoption of the standard no-slip boundary conditions, these
equations provide well-known analytical solutions for simple
geometries \cite[]{Landau,Tuck}. Through these solutions, one
can determine fluidic forces over a body as 
\[F_f(t)= \rho B \omega^2 \Real\{\Gamma(\omega) x(\omega) e^{-i(\omega t+\phi)}\},\] 
where $\rho$ is the fluid density and $B$ is the body volume; 
$\Gamma=m_o(\beta_f+i\gamma_f/\omega)/(\rho B)$ is the
so-called {\it hydrodynamic} function. One of the simplest solutions
of the unsteady NS equations is obtained for an infinite plate
oscillating in a fluid ($L_x=L_z=\infty,L_y=0$), known as the
Stokes' second problem \cite[]{Yakhot,Landau}. For a slender
body having a thin cross section with small but finite width
($0<AR\ll1$) the Newtonian hydrodynamic function can be approximated
by the solution of the Stokes' second problem:
\begin{equation}
\Gamma_{(AR\ll1)}(\omega)= (1+i)~\frac{S}{B}~\sqrt{\frac{\nu}{2\omega}}.
\label{hydro_function_plate}
\end{equation}
Here, $S$ is the surface area in contact with the fluid and
$\nu=\mu/\rho$. When the cross section is not small ($AR\gtrsim 1$),
the common approach has been to study the flow generated by simpler
geometries such as a cylinder with a radius equal to half the
nominal length scale, $r=L_y/2$. For the case of an infinite
cylinder with its axis normal to the $x$-direction, the hydrodynamic
function is \cite[]{Bhiladvala,Sader,Paul}
\begin{equation}
\Gamma_{cyl}(\omega)=1 + \frac{4i}{\sqrt{i\omega r^2/\nu}}~
\frac{\mathrm{K_1}(-i\sqrt{i\omega r^2/\nu})}{\mathrm{K_0}(-i\sqrt{i\omega r^2/\nu})},
\label{hydro_function}
\end{equation}
where $K_0$ and $K_1$ are Bessel's functions of the third kind. The
asymptotic behavior of the Newtonian hydrodynamic function of a
cylinder (\ref{hydro_function}) and that of a rectangular beam are
similar in both limits $\omega r^2/\nu\to 0$ and $\omega r^2/\nu\to
\infty$ \cite[]{Tuck}. For this reason, and only within the Newtonian
regime ($\tau\omega\ll1$), the solution for an oscillating cylinder
has been employed with some degree of accuracy in estimating fluidic
effects on rectangular beams with cross-sections of moderate to
large aspect ratio \cite[]{Bhiladvala,Sader,Paul}. For
very large aspect ratios ($L_x\to 0$), \cite{Sader} formulated
an empirical correction to equation (\ref{hydro_function}) such that
the hydrodynamic function becomes $\Gamma_{(AR\gg1)}=
\Gamma_{cyl}\Omega(\omega L_y^2/\nu)$. Nevertheless, the correction
of \cite{Sader} remains essentially unity $|\Omega-1|={\cal
O}(10^{-1})$ within all regimes studied in this work.

\subsection{Beyond Newtonian hydrodynamics}
\label{beyond} When flow time scales ${\cal T}=1/\omega$ are
of the same order as the relaxation time $\tau$, kinetic effects
become significant and Newtonian hydrodynamic equations
(\ref{eq:nems_N-S}) breakdown. The primary issue encountered beyond
Newtonian regimes ($\tau\omega\gtrsim 1$) is the lack of a robust
hydrodynamic equation governing the flow. To obtain a hydrodynamic
description valid for arbitrary nondimensional frequencies
($0\le\tau\omega\equiv\tau/{\cal T}\le\infty$), one must resort to
kinetic theory representations of the flow. Unfortunately, the
problem of deriving (closed-form) hydrodynamic equations via kinetic
theory, albeit largely studied \cite[see][]{Cercignani,Grad,Chapman,Chen2007},
remains essentially open for arbitrary flow regimes.

Another critical point arising when kinetic effects are no longer
negligible is that of proper boundary conditions for the
hydrodynamic equations at the solid-fluid interface. Hydrodynamic
boundary conditions are determined by a rather complex fluid-solid
interaction in the vicinity of a solid surface. For flows at finite
Knudsen number, the slip boundary condition has been extensively
adopted \cite[]{Cercignani,Lauga,Weng}. According to Maxwell's picture
of slip of a gas over a solid surface, a finite mean-free-path
$\lambda\sim\tau c_s$ leads to an {\it effective} slip velocity
\begin{equation}
[{\bf u}({\bf x}_{w},t)-{\bf U}_w]\bcdot {\bf t}=
\frac{2-\sigma_v}{\sigma_v}~\lambda~
[\bnabla({\bf u}\bcdot{\bf t})\bcdot{\bf n}+\bnabla({\bf u}\bcdot{\bf n})\bcdot{\bf t}]
\label{eq:maxwell_slip}
\end{equation}
to be employed as boundary condition at the coarse-grained
(hydrodynamic) level. Here ${\bf t}$ and ${\bf n}$ are the unit
tangent and normal vectors to a wall located at ${\bf x}_{w}$ and
moving with velocity ${\bf U}_w$; ${\bf u}$ is the fluid velocity.
Meanwhile, $\sigma_v$ is the tangential momentum  accommodation
coefficient of the solid surface. A unit accommodation coefficient
($\sigma_v=1$) represents a situation where all fluid particles are
diffusively scattered after collision with the wall; the opposite
limit ($\sigma_v=0$) corresponds to the case where all such
collisions are specular. First-order Maxwell slip models
(\ref{eq:maxwell_slip}) are accurate for steady flow at small to
moderate Knudsen numbers ($Kn<1$) \cite[]{Weng,Park2003} while
high-order versions have been proposed for unsteady shear flow
\cite[]{Park2003,Hadjiconstantinou2005}. The effective slip in
oscillating shear flows has recently been investigated via kinetic
methods such as lattice Boltzmann-BGK (LBGK) and direct simulation
Monte Carlo (DSMC).  Expected functional shapes have been obtained
for the slip as a function of the Knudsen number $Kn=\lambda/{\cal
L}$ in steady shear flows (${\cal L}\sim |\bnabla u_t|/|u_t|$) or the
nondimensional frequency $\tau\omega=\tau/{\cal T}$ in oscillating
shear flows \cite[]{Park2003,Hadjiconstantinou2005,Colosqui2007}.

\section{Kinetic model of hydrodynamics}
\label{BGK} At room temperature and under ordinary pressure conditions (ranging from low-vacuum to atmospheric pressure), simple gases are composed of a large number of electrically neutral molecules, each with an effective diameter that is negligible
compared to the average intermolecular distance. Under such
conditions the Boltzmann-BGK equation (BE-BGK) is an accurate
kinetic model of the flow \cite[]{Cercignani,Chen2007}. For
monatomic gases in the absence of external force fields ${\bf F}=0$,
the BE-BGK for the evolution of the Boltzmann distribution $f({\bf
x},{\bf v},t)$ in phase space $({\bf x},{\bf v})$ reads:
\begin{equation}
\frac{\partial f}{\partial t}+ {\bf v} \bcdot \bnabla f
=-\frac{f-f^{eq}}{\tau}.
\label{eq:BE-BGK}
\end{equation}
Without loss of generality, we define $\theta=k_B T/m_{gas}=c_s^2$
and adopt a molecular mass $m_{gas}=1$; the equilibrium distribution
can then be expressed as
\begin{equation}
 f^{eq}({\bf x},{\bf v},t)=\frac{\rho}{(2\pi\theta)^{\frac{D}{2}}}\exp\left[-\frac{({\bf v}-{\bf u})^{2}}{2\theta}\right]
\label{eq:feqc}
\end{equation}
where $D$ is the velocity space dimension (${\bf v}=v_k {\bf e}_k;
k=1,D$). Hydrodynamic quantities, like mass density $\rho$, fluid
velocity ${\bf u}$, and energy are obtained as moments of the
distribution function:
\begin{eqnarray}
\int f({\bf x},{\bf v},t) d{\bf v}&=&\rho({\bf x},t), \nonumber\\
\int f({\bf x},{\bf v},t) {\bf v} d{\bf v}&=&\rho{\bf u}({\bf x},t), \\
\int f({\bf x},{\bf v},t) {\bf v}^2 d{\bf v}&=&\rho D \theta({\bf x},t) + \rho {\bf u}^2({\bf x},t). \nonumber
\label{eq:moments}
\end{eqnarray}

{\it Kinetic boundary conditions}. For bounded flows, particular solutions of equation (\ref{eq:BE-BGK}) will require proper boundary conditions. Within the framework of classical kinetic theory we consider the gas as bounded by a perfectly elastic and isothermal surface ($\theta_w=\theta$) located at ${\bf x}_w$ while moving with velocity ${\bf U}_w$. Under this depiction, particles impinging on a solid surface with velocity ${\bf v'}$ acquire a post-collision velocity ${\bf v}$ defined by the scattering kernel $B({\bf v'}\to{\bf v})$. General boundary conditions will then read \cite[]{Cercignani}:
\begin{equation}
|({\bf v}-{\bf U}_w) \bcdot {\bf n}|~f({\bf x}_{w},{\bf v},t)=\!
\int_{({\bf v'}-{\bf U}_w) \bcdot {\bf n}<0} \!\!\!\!\!\!\!\!\!\!\!\!\!\!
 |({\bf v'}-{\bf U}_w) \bcdot {\bf n}|~ B({\bf v'}\to{\bf v}) f({\bf x}_{w},{\bf v'},t) d{\bf v'};
\label{eq:kinetic_bc}
\end{equation}
for $({\bf v}-{\bf U}_w) \bcdot {\bf n}>0$. In this work we implement and assess two different kinetic boundary conditions at the fluid-solid interface; diffuse scattering (DS)
\begin{equation}
f({\bf x}_{w},{\bf v},t)=
\frac{\rho}{(2\pi\theta)^{\frac{D}{2}}}\exp\left[-\frac{({\bf v}-{\bf U}_{w})^{2}}{2\theta}\right]
; ~~ ({\bf v}-{\bf U}_{w})\bcdot{\bf n}>0,
\label{eq:DS}
\end{equation}
and bounce-back (BB)
\begin{equation}
f({\bf x}_{w},{\bf v},t)=f({\bf x}_{w},-{\bf v}+2{\bf U_{w}},t); ~~ ({\bf v}-{\bf U}_{w})\bcdot{\bf n}>0,
\label{eq:BB}
\end{equation}
where impinging particles [$({\bf v}-{\bf U}_{w})\bcdot{\bf n}<0$] are reflected back  with the same relative speed and angle of incidence. While a diffuse-scattering (DS) kernel yields hydrodynamic slip in agreement with the Maxwell slip model (\ref{eq:maxwell_slip}) for a fully accommodating surface ($\sigma_v=1$), the bounce-back model renders no-slip at the wall for all flow regimes (i.e. the surface has negative accommodation coefficient $\sigma_v=-2$). The validity range of each model for the solid-fluid interaction will be assessed when comparing LBGK simulation using both DS and BB schemes against experimental results.

\subsection{Free-molecule hydrodynamics}
The nondimensional frequency $\tau\omega = 2 \pi\lambda/{\cal
L}_{FM}$ is proportional to the ratio of the (equilibrium) mean free
path $\lambda=\tau c_s$ to the average distance ${\cal L}_{FM}=c_s
2\pi/\omega$ traveled by a particle during one oscillation period.
Hence, in the limit $\tau\omega \to \infty$, the kinetic dynamics becomes
practically collisionless and free-molecular flow approximations are
applicable. Let us now analyze a $L_x \times L_y$ rectangular
section in free molecular flow moving in the $x$-direction at
velocity ${\bf u}=u {\bf i}$ such that $M=U_w/\sqrt{\theta}\ll1$. In the case of diffusive wall scattering (\ref{eq:DS}) the net $x$-force per unit length is \cite[]{Bird}
\begin{equation}
F_{f}^{DS}(t)=-\left(\sqrt{\frac{8}{\pi}}L_y+\sqrt{\frac{2}{\pi}} L_x\right)
\rho \sqrt{\theta}~ u.
\label{eq:free-molecular}
\end{equation}
For the bounce-back model (\ref{eq:BB}), which renders no
hydrodynamic slip, one has $F_{f}^{BB}= 2 F_{f}^{DS}$. Similar
expressions can be obtained for other models of the gas-surface
interaction, e.g.  $F_{f}^{SP}(t)=-\left(\sqrt{\frac{32}{\pi}}L_y
\right)\rho \sqrt{\theta}~ u$ for specular reflection models
rendering no shear stress. Clearly, in free molecular flow
($\tau\omega\to\infty$), there is no fluidic inertia ($\beta_f\to0$)
and hydrodynamic forces only have a dissipative component
($F_{f}=f_d~u$).

\section{Lattice Boltzmann BGK simulation}
\label{sec:LBGK} 
The method in this work falls in the
class of high-order LBGK models originally formulated by \cite{Shan2006,Shan2007} and investigated by \cite{Colosqui2009} for isothermal unidirectional flow in
non-Newtonian regimes. Discretization of velocity space in a finite
number of lattice velocities $\{ {\bf v}_i;~i=1,Q\}$ allows one to
reduce the problem of solving equation  (\ref{eq:BE-BGK}) to that of
solving a set of lattice Boltzmann-BGK (LBGK) equations
\begin{equation}
\frac{\partial f_i}{\partial t}+{\bf v}_i \bcdot \bnabla f_i = - \frac{f_i -
 f_i^{eq}}{\tau}
\label{LBGK}
\end{equation}
\begin{equation}
~~~~~~~~~
f^{eq}_i = w_i \rho
[ 1 + {\textstyle \frac{1}{\theta}}({\bf v}_i \bcdot {\bf u})
+{\textstyle \frac{1}{2\theta^2}}({\bf v}_i \bcdot {\bf u})^2
-{\textstyle \frac{1}{2\theta}}{\bf u}^2]~~~~~i=1,Q
\label{eq:feq_h2}
\end{equation}
governing the evolution of each lattice population $f_i({\bf
x},t)=w_i f({\bf x},{\bf v}_i,t)$ in configuration space. The set of
LBGK equations (\ref{LBGK}-\ref{eq:feq_h2}) is formally derived by
projecting the continuum Boltzmann-BGK equation
(\ref{eq:BE-BGK}-\ref{eq:feqc}) onto the Hilbert space ${\mathrm
H}^{2}$ spanned by the orthonormal basis of Hermite polynomials up
to second order. A Gauss-Hermite quadrature formula determines the
lattice velocities ${\bf v}_i$, i.e. integration points, and their
associated weights $w_i$. A quadrature rule with algebraic degree of
precision $d\ge 4$ permits the exact numerical integration of the
three leading moments (\ref{eq:moments}) of the continuum
distribution $f$:
\begin{equation}
\rho({\bf x},t)= \sum_{i=1}^{Q} f_i({\bf x},t),~~
\rho {\bf u}({\bf x},t)= \sum_{i=1}^{Q} f_i({\bf x},t) {\bf v}_i,~~
\rho (u^2 +D \theta)({\bf x},t)
=\sum_{i=1}^{Q} f_i({\bf x},t) {\bf v}_i^2.
\label{eq:MME_lbgk}
\end{equation}
The particular lattice employed for the present simulations is the
D2Q37 \cite[]{Shan2007} (i.e. velocity space dimension $D\!=\!2$,
number of lattice velocities $Q\!=\!37$); the weights and velocity
abscissae of the lattice are included in the appendix. After lattice
discretization of configuration space (${\bf \Delta x}_i={\bf v}_i
\Delta t$) numerical procedures to solve the LBGK equations
(\ref{LBGK}) advance in two steps: advection and collision. During
the advection step the streaming of lattice populations $\{ f_i;
i=1,Q \}$ is performed along the lattice directions:
\begin{equation}
f_{i}^{adv}({\bf x}, t)=f_{i}({\bf x}-{\bf v}_i \Delta t, t).
\label{advection}
\end{equation}
At the end of the advection step, mass, momentum, and energy (\ref{eq:MME_lbgk}) are computed using $f_i^{adv}$ allowing the explicit evaluation of equilibrium distributions (\ref{eq:feq_h2}). Then the BGK ansatz is applied at the collision step:
\begin{equation}
f_i({\bf x}+{\bf v}_i,t+\Delta t)=
f_i^{eq} + \left[ 1 - \frac{\Delta t}{\tau} \right] \widehat{f_{i}}^{ne}.
\label{regularization}
\end{equation}
The projected, or {\it dealiased}, nonequilibrium component \cite[]{Zhang}
\begin{equation}
\widehat{f}_i^{ne}= \frac{w_i}{2 \theta^2} ({v_{\alpha}}_{i}{v_{\beta}}_{i} -\theta \delta_{\alpha\beta}) \sum_{j=1}^{Q}(f_j^{adv}-f_j^{eq}) {v_{\alpha}}_{j}{v_{\beta}}_{j} ~~~~~ \alpha,\beta=1,D~~i=1,Q
\label{eq:fne}
\end{equation}
ensures that the post-collision distribution $f_i$ is strictly expressed by a linear combination of up to second-order Hermite polynomials, i.e $f_i \in {\mathrm H}^{2}$. The {\it regularization} procedure \cite[]{Zhang} specified in equations (\ref{eq:fne})-({\ref{regularization}) eliminates significant numerical errors due to lattice orientation anisotropy and {\it aliasing} effects that are reported for the standard LBGK algorithm in strong non-equilibrium conditions \cite[]{Colosqui2009,Zhang}.

\section{Results}
\label{results} The quantitative analysis of fluidic damping on
devices with different sizes and structural features is performed by
employing the nondimensional quantity:
\begin{equation}
\overline{\gamma}_f(\tau\omega,AR)=
\frac{\gamma_f}{\rho \sqrt{\frac{\theta}{2}}~ \frac{S}{m_o}}.
\label{eq:nems_W_tw}
\end{equation}
Here, $S/m_o$ is the ratio of wet area to modal mass reported in table~\ref{tab:nems}. After normalization by the fluid mass density $\rho$ and speed of sound $c_s=\sqrt{\theta}$, the dimensionless damping $\overline{\gamma}_f$ (\ref{eq:nems_W_tw}) is solely determined by the nondimensional frequency $\tau\omega$ and hydrodynamic shape characterized by the aspect ratio $AR=L_y/L_x$. Based on reported data in table~\ref{tab:nems}, we adopt $AR\simeq 0$ for the quartz resonator and $AR\simeq$1, 10 \& 26 for the beams. Experimentally measured values of the total damping are presented as open symbols in figure~\ref{fig:nems_3}. These are determined from measurements of the resonant response (\ref{eq:nems_uw}) as a function of pressure as discussed above.  As observed in figure~\ref{fig:nems_3}, experimental values of the density-normalized damping $\overline{\gamma}_f$ (\ref{eq:nems_W_tw}) exhibit an
increasing dispersion in the upper limit of dimensionless frequency variation $\tau\omega\to\infty$ where $\gamma_f \to 0$. This is due to the fact that the measured fluidic effect becomes smaller and smaller as the pressure is lowered and the measured signal is dominated by the finite error $\sim 5\%$  present in all the
experimental data. This is discussed in more detail in \cite{Karabacak2007}.

{\it Lattice Boltzmann-BGK (LBGK) simulation}. Simulations
are performed with the D2Q37-H2 model described in Sec.~\ref{sec:LBGK}. It has been demonstrated that high-order LBGK models such as D2Q37-H2 with a {\it regularization} procedure yield excellent agreement with extended hydrodynamic descriptions derived
for isothermal and unidirectional flows in both Newtonian and  non-Newtonian regimes \cite[]{Colosqui2009,Zhang}. In all simulations, $\tau=\mu/p\equiv\nu/\theta$ is readily determined from the gas properties at thermodynamic equilibrium, while the employed Mach number is very low $M=U_o/\sqrt{\theta}\le 0.01$. Simulation results on different resonators obtained by the D2Q37-H2 model for 
$\tau\omega$=0.001, 0.01, 0.1, 0.5, 1, and 10 are compared against
experimental data and available analytical expressions for Newtonian
and free-molecular flow in figure~\ref{fig:nems_3}. The employed
LBGK models yield a remarkable agreement with experimental
measurements over a wide range of nondimensional frequency
$0.001\le\tau\omega\le10$ for different device geometries and
dimensions (table~\ref{tab:nems}). On the other hand, Newtonian
fluid approximations such as equation (\ref{hydro_function_plate})
for the quartz disk ($AR\simeq0$) and (\ref{hydro_function}) for the
beams ($AR\ge1$) \cite[]{Bhiladvala,Sader,Paul}  give
acceptable agreement only within the low frequency limit
$\tau\omega<0.1$. It is important to remark that all Newtonian
predictions largely overestimate the fluidic dissipation in the
frequency range $\tau\omega>0.1$. In the high-frequency limit
$\tau\omega\gg 1$, only kinetic approaches such as LBGK simulation
and the free-molecular flow solution given by equation (\ref{eq:free-molecular})
are in good agreement with experiment.

{\it Fluid-solid interaction and boundary schemes}. Owing
to the small amplitude of oscillation ($|x(\omega)|\ll L_x$), the
solid boundary can be assumed to remain static. Only the boundary
velocity ${\bf U_{w}}=U_o \sin(\omega t) {\bf i}$ varies in time
with velocity amplitude $U_o<\omega \Delta x$ such that the
displacement amplitude is smaller than the distance between
neighboring lattice nodes. Two boundary schemes are employed in LBGK
simulation when modeling the moving walls: the diffuse-scattering
scheme (DS) and the bounce-back scheme (BB) that were explained in
Sec. \ref{BGK}. While the BB scheme (\ref{eq:BB}) renders no-slip as
hydrodynamic boundary condition, the DS model (\ref{eq:DS}) produces
an effective slip velocity between the body surface and the fluid
immediately adjacent to it. Experimental results in figure~\ref{fig:nems_3} show that numerical schemes rendering no-slip at
the wall overpredict the fluidic dissipation when $\tau\omega \ge
0.1$. On the other hand, the diffuse-scattering scheme which models
the resonator surface as fully accommodating ($\sigma_v=1$) yields a
close agreement with experimental data in the entire studied range
$0.001\le \tau\omega \le 10$.

\begin{figure}
\centerline{\includegraphics[angle=0,scale=0.4]{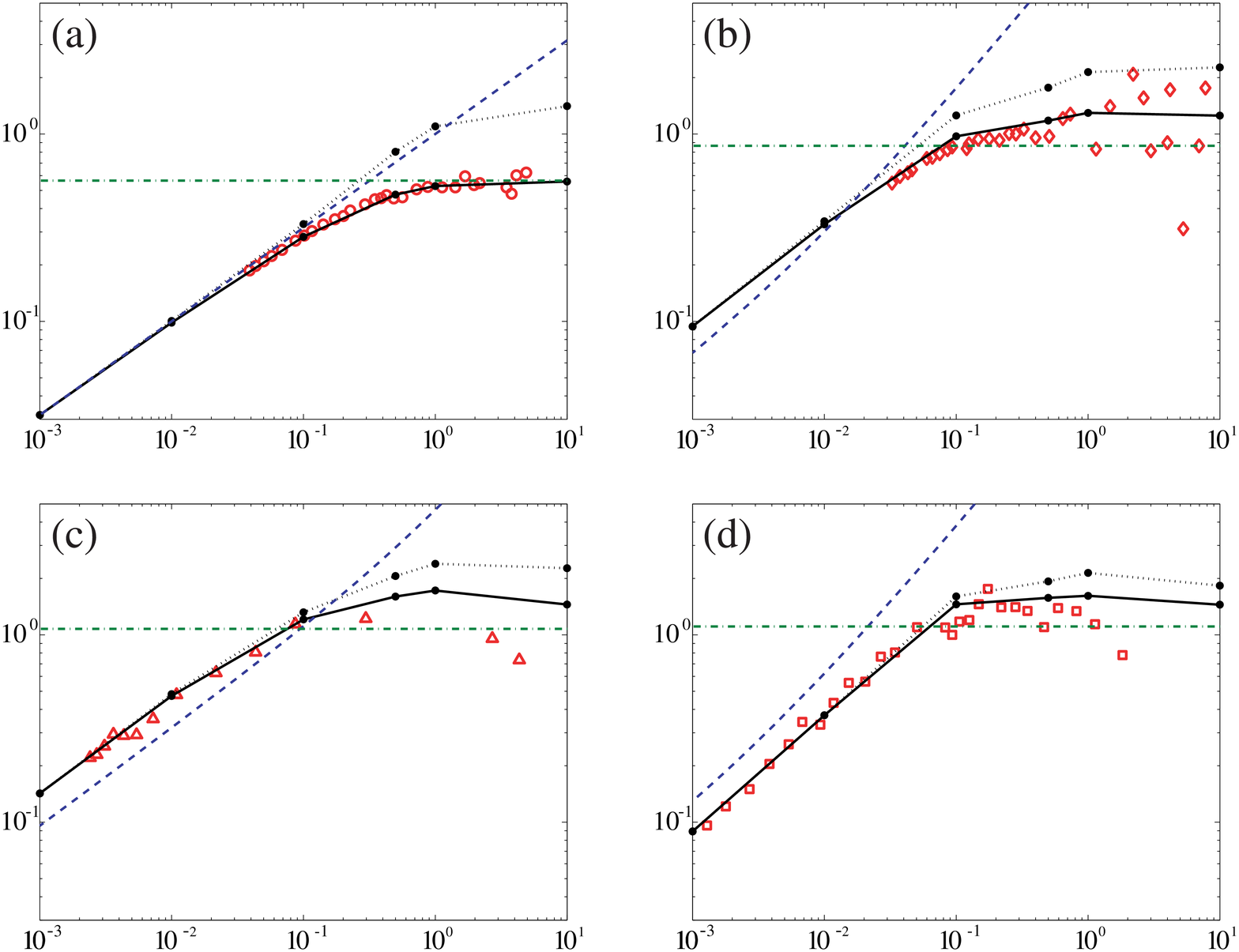}}
\caption{Nondimensional fluid damping:
$\overline{\gamma}_f(\tau\omega,AR)$ vs. $\tau\omega$. (a) $\omega_o/2\pi\!=\!32.7$ MHz, $AR\!=\!0$. (b) $\omega_o/2\pi\!=\!24.2$ MHz, $AR\!=\!1$. (c) $\omega_o/2\pi\!=\!1.97$ MHz,
$AR\!=\!10$. (d) $\omega_o/2\pi\!=\!0.078$ MHz, $AR\!=\!26$.
Open symbols (red): experimental data.
Closed dots (black): LBGK simulation $\tau\omega$=0.001, 0.01, 0.1, 0.5, 1, and 10;
solid line  (black): D2Q37-H2 DS;
dotted line (black): D2Q37-H2 BB.
Dashed line (blue): Newtonian fluid approximations.
Dashed-dotted line (green): free molecular flow (DS).}
\label{fig:nems_3}
\end{figure}

{\it Quality factor and fluidic effects}. Since fluidic
inertia is very small ($\beta_f\ll 1$) for the studied flow
conditions the quality factor of the fluid-immersed device is
\begin{equation}
Q=Q_o\frac{1}{1+\rho \sqrt{\frac{\theta}{2}}~\frac{S}{m_o\gamma_o}\overline{\gamma}_f(\tau\omega,AR)}.
\end{equation}
Evidently, dissipative effects quantified by the structural dissipation $\gamma_o=\omega_o/Q_o$ and fluidic damping $\gamma_f=\rho \sqrt{\frac{\theta}{2}}~\frac{S}{m_o}\overline{\gamma}_f$ determine the device performance. The quality factors reported for the four studied devices are compared against numerical predictions in figure~\ref{fig:nems_7}. We observe that lattice Boltzmann-BGK simulation
(model D2Q37-H2 with DS wall treatment) is in close agreement with
experimental data obtained for different cross-sections ($0\le AR
\le 26 $) in a wide range of operation conditions ($0.1\le p\le
1000$ Torr, $0.078\le\omega_o\le24$~MHz).

\begin{figure}
\centerline{\includegraphics[angle=0,scale=0.5]{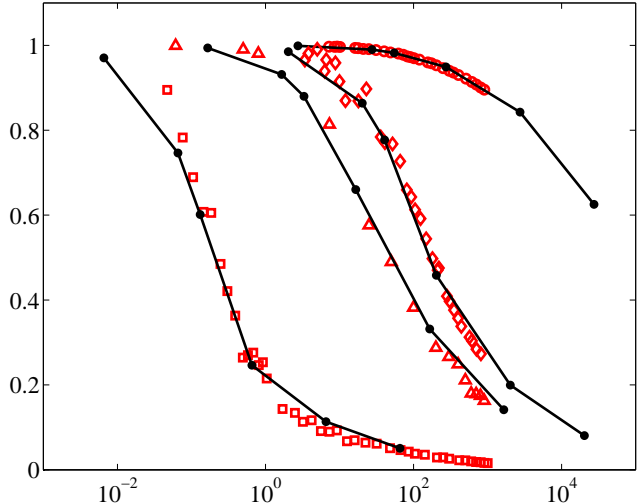}}
\caption{Quality Factor: {\large $\frac{Q}{Q_o}=\frac{\gamma_o}{\gamma_o+\gamma_f}$} vs. $p$~[Torr]. 
Open symbols (red): experimental data; ({\Large $\circ$}) $\omega_o/2\pi\!=\!32.7$ MHz, $AR\!=\!0$; ({\Large $\diamond$}) $\omega_o/2\pi\!=\!24.2$ MHz, $AR\!=\!1$;
($\bigtriangleup$) $\omega_o/2\pi\!=\!1.97$ MHz, $AR\!=\!10$;
($\square$) $\omega_o/2\pi\!=\!0.078$ MHz, $AR\!=\!26$.
Closed dots (black): LBGK simulation (D2Q37-H2 DS) $\tau\omega$=0.001, 0.01, 0.1, 0.5, 1, and 10.}
\label{fig:nems_7}
\end{figure}

\section{Conclusions}
\label{conclusions} Experimental measurements of fluidic effects on
diverse electromechanical resonators have been compared against
available analytical approaches, i.e. the Newtonian flow  and
free-molecular flow approximations, as well as against kinetic-based
simulations presented in Sec.~\ref{sec:LBGK}. Clearly, Newtonian and free-molecular flow models fail to describe transitional flow in the region $0.1\le\tau\omega\le10$. On the other hand, lattice Boltzmann-BGK (LBGK) simulation accounting for specific geometrical features accurately represents the fluidic damping $\gamma_f$ in all studied regimes: Newtonian ($\tau\omega\le 0.1$), transitional ($0.1\le \tau\omega\le10$), and free-molecular flow ($\tau\omega\ge10$).

{\it Viscoelastic dynamics in high-frequency flows}. The
invalidity of Newtonian approaches as $\tau\omega \to \infty$ is not
only due to surface effects, which might be absorbed by proper
hydrodynamic boundary conditions, but also due to the qualitatively
different fluid dynamics in the bulk \cite[]{Yakhot,Colosqui2009}.
Kinetic-based (LBGK) simulation precisely reproduces the
experimentally observed (figure~\ref{fig:nems_3}) saturation of
density-normalized dissipation ($\overline{\gamma}_f \to const$) in
the high-frequency limit $\tau\omega\to\infty$. This remarkable
phenomenon involves a gradual transition from viscous to
viscoelastic to purely elastic flow of a simple monatomic gas that
has been reported by previous theoretical \cite[]{Yakhot,Colosqui2009}
and experimental \cite[]{Karabacak2007} studies. The viscoelastic
response of simple gases in the high-frequency limit is a well-known
phenomenon within the realm of transport theory and statistical
physics \cite[]{Evans}; diffusion processes in nonequilibrium systems
can only be established after a finite time ${\cal T}_{D}\sim
\tau$ of the order of the relaxation time. In the short-time limit
$t<\tau$, where diffusion effects are still weak and transport
coefficients such as shear viscosity become frequency dependent
\cite[]{Evans}, one observes an ensuing decay in the dissipation of
fluid momentum and energy.

{\it Near-wall phenomena}. Kinetic effects can never be
neglected within the so-called Knudsen layer, e.g. at distances from
the solid boundary that are smaller than one mean free path. Thus,
kinetic parameters, such as the relaxation time and mean free path,
are not easily determined in the near-wall region where gas-solid
interaction is significant. In consequence, within a mean free path
from the wall, Newtonian fluid models for the stress break down and
NS equations must actually be  applied immediately outside the
Knudsen layer. The concept of having effective slip as proper
boundary condition for hydrodynamic (coarse-grained) equations must
be understood within this context. Our kinetic model of the flow
based on the Boltzmann-BGK is rather simple and does not accurately
represent the Knudsen layer; it relies on a single constant
relaxation time $\tau=\mu/p$ and kinetic boundary conditions
(\ref{eq:kinetic_bc}) determined by surface scattering kernels
$B({\bf v'}\to{\bf v})$ for a perfectly elastic and isothermal
surface. Nevertheless, the kinetic model in this work accurately
predicts hydrodynamic effects such as fluid resistance and mean
energy dissipation via adoption of a Maxwell scattering kernel with
surface  accommodation $\sigma_{v}=1$ (i.e. the DS scheme explained
in Sec. \ref{BGK}). The net effect of the studied gas-surface
interactions can be assessed by comparing results (figure~\ref{fig:nems_3}) from DS (slip) and BB (no-slip) schemes in the
range $0.001\le\tau\omega\le10$ for different cross-sections $0\le
AR\le26$. The reduction in fluid damping solely due to effective
slip is found in the interval $0-0.4$.

{\it Resonator performance beyond Newtonian regime}. The
decay of the energy dissipated by the fluid as $\tau\omega \to
\infty$ has beneficial effects on the resonator performance. Under
relevant experimental conditions, fluidic damping largely dominates
over structural dissipation, $\gamma_f\gg\gamma_o$, and thus,
$Q\approx \omega/\gamma_f$. In such conditions the quality factor
will actually increase linearly with the operation frequency,
$Q\propto \omega$ for $\tau\omega>1$, instead of the square root
dependence, $Q\propto \sqrt{\omega}$, observed for Newtonian flow
$\tau\omega\ll 1$. Therefore, it is  advantageous to operate the
resonator at a frequency $\tau\omega>1$ well within the viscoelastic
regime. This could be accomplished either by  increasing the
resonance frequency or by decreasing the effective relaxation time
of the fluid through less trivial mechanisms, e.g. through polymer
addition or foams for water. Other strategies that can potentially
improve the device performance include modifying the cross-sectional
shape and surface properties. As observed in figure~\ref{fig:nems_3}
for the entire range $0.001\le\tau\omega\le 10$, bodies with small
aspect ratios $AR=L_y/L_x\ll 1$ generate less fluidic damping at the
same operation frequency and surrounding gas conditions. On the
other hand, the employment of superhydrophobic coatings for
resonators in water can further increase the effective hydrodynamic
slip with a subsequent reduction of the resistance forces.

{\it Lattice Boltzmann-BGK simulation for N/MEMS Hydrodynamic}. The qualitative and quantitative agreement between
our LBGK simulations and experimental data over a wide range of
pressure $0.1\le p \le 1000$ Torr and frequency variation $0.001\le
\tau\omega \le 10$ constitutes a remarkable achievement for the
kinetic methodology applied in this work. The precise determination
of fluid forces and, thus, device quality factors for diverse
geometrical configurations in widely different operation regimes is
of fundamental importance in advancing the design of future N/MEMS
devices.

\section{Acknowledgments}
The authors acknowledge Dr. Hudong Chen and Xiaowen Shan from EXA
corporation for their support in the development of the employed
numerical tools. This work was funded by the National 
Science Foundation (NSF) under Grant No. CBET-0755927.

\appendix
\section{Appendix: The LBGK model D2Q37}
\label{app:LBGK}
Velocity abscissae and weights of the D2Q37 lattice model \cite[]{Shan2006,Shan2007} is presented in table~\ref{tab:d2q37}
\begin{table}
\begin{center}
\begin{minipage}{7.5cm}
\begin{tabular}{@{}ccc@{}}
 ${\bf v}_i/c$ \footnote{Lattice Speed: $c$ =  1.19697977039307 $\sqrt{\theta}$} & states & $w_i$ \\[0.5ex]
$(0,0)$                 & 1 & $0.233150669132352000$ \\[0.5ex]
$(1,0)$            & 4 & $0.107306091542219000$   \\[0.5ex]
$(\pm 1,\pm 1)$         & 4 & $0.057667859888794800$  \\[0.5ex]
$(2,0)$            & 4 & $0.014208216158450700$   \\[0.5ex]
$(\pm 2,\pm 2)$         & 4 & $0.001011937592673570$  \\[0.5ex]
$(3,0)$            & 4 & $0.000245301027757717$   \\[0.5ex]
$(\pm 1,\pm 2)$        & 8 & $0.005353049000513770$   \\[0.5ex]
$(\pm 1,\pm 3)$        & 8 & $0.000283414252994198$   \\[0.5ex]
\end{tabular}
\end{minipage}
\end{center}
\caption{Model Parameters D2Q37}
\label{tab:d2q37}
\end{table}

\bibliographystyle{plain}

\end{document}